\documentclass[12pt,preprint]{aastex}

\shorttitle{TeV Observations of the A426 and A2029 Clusters}
\shortauthors{Perkins et al.}

\begin{document}

\title{TeV Gamma-Ray Observations of the Perseus and Abell 2029 Galaxy Clusters}

\author{J.S. Perkins\altaffilmark{1},
  H. M. Badran\altaffilmark{2}, 
  G. Blaylock\altaffilmark{3},
  S. M. Bradbury\altaffilmark{4},
  P. Cogan\altaffilmark{5},
  Y.C.K. Chow\altaffilmark{6},
  W. Cui\altaffilmark{7},
  M. K. Daniel\altaffilmark{5},
  A. D. Falcone\altaffilmark{8},
  S.J. Fegan\altaffilmark{6}, 
  J.P. Finley\altaffilmark{7},
  P. Fortin\altaffilmark{9},
  L. F. Fortson\altaffilmark{10},
  G. H. Gillanders\altaffilmark{11},
  K.J. Gutierrez\altaffilmark{1},
  J. Grube\altaffilmark{4},
  J. Hall\altaffilmark{12},
  D. Hanna\altaffilmark{13}, 
  J. Holder\altaffilmark{4},
  D. Horan\altaffilmark{14}, 
  S.B. Hughes\altaffilmark{1},
  G. E. Kenny\altaffilmark{11},
  M. Kertzman\altaffilmark{15},
  D. B. Kieda\altaffilmark{12},
  J. Kildea\altaffilmark{13},
  K. Kosack\altaffilmark{1},
  H. Krawczynski\altaffilmark{1}, 
  F. Krennrich\altaffilmark{16},
  M.J. Lang\altaffilmark{11},
  S. LeBohec\altaffilmark{12},
  G. Maier\altaffilmark{4},
  P. Moriarty\altaffilmark{17},
  R.A. Ong\altaffilmark{6},
  M. Pohl\altaffilmark{16}, 
  K. Ragan\altaffilmark{13},
  P.F. Rebillot\altaffilmark{1},
  G.H. Sembroski\altaffilmark{7}, 
  D. Steele\altaffilmark{10},
  S.P Swordy\altaffilmark{18},
  L. Valcarcel\altaffilmark{13}, 
  V. V. Vassiliev\altaffilmark{6},
  S.P. Wakely\altaffilmark{18}, 
  T. C. Weekes\altaffilmark{14},
  D. A. Williams\altaffilmark{19},
  (The VERITAS Collaboration)}

\altaffiltext{1}{Department of Physics, Washington University in
  St. Louis, St. Louis, MO 63130, USA}

\altaffiltext{2}{Physics Department, Tanta University, Tanta, Egypt}

\altaffiltext{3}{Department of Physics, University of Massachussetts,
Amherst, MA 01003-4525, USA}

\altaffiltext{4}{School of Physics and Astronomy, University of Leeds,
Leeds, LS2 9JT, UK}

\altaffiltext{5}{School of Physics, University College Dublin,
Belfield, Dublin 4, Ireland}

\altaffiltext{6}{Department of Physics and Astronomy, University of
California, Los Angeles, CA 90095, USA}

\altaffiltext{7}{Department of Physics, Purdue University, West
  Lafayette, IN 47907, USA}

\altaffiltext{8}{Department of Astronomy and Astrophysics, Penn State
University, University Park, PA 16802, USA}

\altaffiltext{9}{Department of Physics and Astronomy, Barnard College,
Columbia University, NY 10027}

\altaffiltext{10}{Astronomy Department, Adler Planetarium and Astronomy
Museum, Chicago, IL 60605, USA}

\altaffiltext{11}{Physics Department, National University of Ireland,
Galway, Ireland}

\altaffiltext{12}{Physics Department, University of Utah, Salt Lake
City, UT 84112, USA}

\altaffiltext{13}{Physics Department, McGill University, Montreal, QC
H3A 2T8, Canada}

\altaffiltext{14}{Fred Lawrence Whipple Observatory,
  Harvard-Smithsonian Center for Astrophysics, Amado, AZ 85645, USA}

\altaffiltext{15}{Department of Physics and Astronomy, DePauw
  University, Greencastle, IN 46135-0037, USA}

\altaffiltext{16}{Department of Physics and Astronomy, Iowa State
University, Ames, IA 50011, USA}

\altaffiltext{17}{Department of Physical and Life Sciences,
Galway-Mayo Institute of Technology, Dublin Road, Galway, Ireland}

\altaffiltext{18}{Enrico Fermi Institute, University of Chicago,
Chicago, IL 60637, USA}

\altaffiltext{19}{Santa Cruz Institute for Particle Physics and
Department of Physics, University of California, Santa Cruz, CA 95064,
USA}

\email{corresponding authors: Jeremy Perkins
  <jperkins@physics.wustl.edu> and Henric Krawczynski
  <krawcz@wuphys.wustl.edu>}

\begin{abstract}
  Galaxy clusters might be sources of TeV gamma rays emitted by
  high-energy protons and electrons accelerated by large scale
  structure formation shocks, galactic winds, or active galactic
  nuclei.  Furthermore, gamma rays may be produced in dark matter
  particle annihilation processes at the cluster cores.  We report on
  observations of the galaxy clusters Perseus and Abell 2029 using the
  10 m Whipple Cherenkov telescope during the 2003-2004 and 2004-2005
  observing seasons.  We apply a two-dimensional analysis technique to
  scrutinize the clusters for TeV emission.  In this paper we first
  determine flux upper limits on TeV gamma-ray emission from point
  sources within the clusters.  Second, we derive upper limits on the
  extended cluster emission.  We subsequently compare the flux upper
  limits with {\it EGRET} upper limits at 100 MeV and theoretical
  models. Assuming that the gamma-ray surface brightness profile
  mimics that of the thermal X-ray emission and that the spectrum of
  cluster cosmic rays extends all the way from thermal energies to
  multi-TeV energies with a differential spectral index of -2.1, our
  results imply that the cosmic ray proton energy density is less than
  7.9\% of the thermal energy density for the Perseus cluster.
\end{abstract}


\section{Introduction}

As our Universe evolves and structure forms on increasingly larger
scales, the gravitational energy of matter is constantly converted
into random kinetic energy of cosmic gas. In galaxy clusters,
collisionless structure formation shocks are thought to be the main
agents responsible for heating the inter-cluster medium (ICM) to
temperatures of $k_{\rm B}$T$\simeq$10~keV.  Through this and other
processes, gravitational energy is converted into the random kinetic
energy of non-thermal baryons (protons) and leptons (electrons).
Galactic winds \citep{voelk.1999} and re-acceleration of mildly
relativistic particles injected into the ICM by powerful cluster
members \citep{ensslin.1998} may accelerate additional particles to
non-thermal energies. Using galactic cosmic rays (CR) as a yard stick,
one expects that the energy density of cosmic ray protons (CRp)
dominates over that of cosmic ray electrons (CRe) by approximately two
orders of magnitude, and may be comparable to that of thermal
particles and the ICM magnetic field. CRp can diffusively escape
clusters only on time scales much longer than the Hubble
time. Therefore they accumulate over the entire formation history
\citep{voelk.1999}. CRe lose their energy by emitting synchrotron,
Bremsstrahlung, and inverse Compton emission on much shorter time
scales. For ICM magnetic fields on the order of $B\simeq1\mu$G,
synchrotron and inverse Compton emission losses alone cool CRe of
energy $E=1$~TeV on a timescale
\begin{equation}
\label{tau}
\tau_s = \left(\frac{4}{3}\sigma_Tc\frac{B'^2}{8\pi m_e c^2}\gamma_e\right)^{-1}
\end{equation}
where $\sigma_T$ is the Thomson cross section,
    $B'=\sqrt{B^2+B_{\rm{CMB}}^2}$ and $B_{\rm{CMB}} =
    3.25(1+z)^2\mu$G; for the clusters considered here, $z \ll 1$ and
    $\tau_s \approx 10^6$~years.

There is good observational evidence nonthermal electrons in galaxy
clusters. For a number of clusters, diffuse synchrotron radio halos
and/or radio relic sources have been detected \citep{giovannini.1993,
giovannini.1999, giovannini.2000, kempner.2001, feretti.2003}.  For
some clusters, an excess of Extreme Ultra-Violet (EUV) and/or hard
X-ray radiation over that expected from the thermal X-ray emitting ICM
has been observed \citep{bowyer.1998, lieu.1999, rephaeli.1999,
fusco-femiano.2004}. The excess radiation originates most likely as
inverse Compton emission from CRe scattering cosmic microwave
background photons \citep{lieu.1996, ensslin.1998, blasi.1999,
fusco-femiano.1999}.

The detection of gamma-ray emission from galaxy clusters would make it
possible to measure the energy density of non-thermal particles.  The
density and energy density of the thermal ICM can be derived from
imaging-spectroscopy observations made with such satellites as
\textit{Chandra} and \textit{XMM-Newton} \citep{krawczynski.2002,
markevitch.1998, donahue.2004}.  The density and energy spectra of the
non-thermal protons could be computed from the detected gamma-ray
emission once the density of the thermal ICM is known
\citep{pfrommer.2004}. Gamma rays can originate as inverse Compton and
Bremsstrahlung emission from CRe and as $\pi^0\rightarrow
\gamma\gamma$ emission from hadronic interactions of CRp with thermal
target material. If successful measurements of the gamma-ray fluxes
from several galaxy clusters were obtained, one could explore the
correlation of the CRp luminosity with cluster mass, temperature, and
redshift, and draw conclusions about how the clusters grew to their
observed size.  If CRp indeed contribute noticeably to the pressure of
the ICM, the measurements of the CRp energy density would allow
improvement on the estimates of the cluster mass based on X-ray data,
and thus improve estimates of the universal baryon fraction.  If CR
provide pressure support to the ICM, they would inhibit star formation
to some extent as they do not cool radiatively like the thermal X-ray
emitting gas. Furthermore, low energy cosmic ray ions might provide a
source of heating the thermal gas \citep{rephaeli.1977}.

In addition to a CR origin, annihilating dark matter may also emit
gamma rays. The intensity of the radiation depends on the nature of
dark matter, the annihilation cross sections, and the dark matter
density profile close to the core of the cluster,
e.g. \cite{bergstrom.1998}.  While MeV observations are ideally suited
for detecting the emission from the bulk of the non-thermal particles,
TeV gamma-ray observations of cluster energy spectra and radial
emission profiles would allow us to disentangle the various components
that contribute to the emission.

At MeV energies, various authors have searched for cluster emission
based on the data from the {\it EGRET} {\it (Energetic Gamma Ray
Experiment Telescope)} detector on board the {\it Compton Gamma-Ray
Observatory}.  Three studies revealed evidence at a significance level
of approximately three standard deviations: \cite{colafrancesco.2001}
and \cite{kawasaki.2002} reported an association between Abell
clusters and unidentified gamma-ray point sources from the third
catalog of the {\it EGRET} experiment; \cite{scharf.2002} found
gamma-ray emission from Abell clusters by stacking the {\it EGRET}
data of 447 galaxy clusters.  However, analyzing the data from 58
galaxy clusters, \cite{reimer.2003} do not confirm a detection, and
give an upper limit that is inconsistent with the mean flux reported
by \cite{scharf.2002}. In the TeV energy range, \cite{fegan.2005}
reported marginal evidence for emission from Abell~1758 in the field
of view of 3EG~J1337~+5029.

In this paper we report on a search for TeV gamma-ray emission from
the Perseus and Abell 2029 galaxy clusters with the Whipple 10~m
Cherenkov telescope.  We selected both clusters based on their
proximity and high masses: Perseus ($z\,=$ 0.0179) is at a distance of
75 Mpc from us and has a total mass of 4$\times10^{14}$~M$_\odot$;
Abell 2029 ($z\,=$ 0.0775) is 300 Mpc away from us and its total mass has
been estimated to be 5$\times10^{14}$~M$_\odot$ \citep{girardi.1998}.

The search described below assumes that the high energy (HE) surface
brightness mimics the X-ray surface brightness, and focuses on the
detection of gamma rays from within 0.8 degrees from the cluster
center.  There are several possibilities relating the thermal and
non-thermal particles within clusters. From general considerations,
\cite{voelk.1999} assume that the non-thermal particles carry a
certain fraction of the energy density of the ICM. One of the aims of
VHE astronomy is to constrain this fraction. Indeed, we do know the
CRp energy density in the Interstellar Medium (ISM) of the Milky Way
galaxy. In this case it turns out that the CRp energy density is
comparable to the energy density of the thermal ISM, the energy
density of the interstellar magnetic field and the energy density of
star light. If non-thermal particles in clusters indeed carry a
certain fraction of the energy density of the ICM, the HE surface
brightness would mimic that of the thermal X-ray emission.  In another
line of argument, one may assume that powerful cluster members
(i.e. radio sources) are the dominant source of non-thermal particles
in the ICM; also in this case we would expect that CRp accumulate at
the cluster cores where usually the most powerful radio galaxies are
found \citep{pfrommer.2004}. \cite{ryu.2003, kang.2005} performed numerical
calculations to estimate the energy density of CRp by large scale
structure formation shocks.  The simulations indicate that strong
shocks form preferentially in the cluster periphery.  Accordingly,
most CRp would be accelerated in the outskirts of the clusters and
would only slowly be transported to the cluster core by bulk plasma
motion (e.g. following cluster merger). The main conclusion of this
discussion is that the CRp distribution in galaxy clusters is
uncertain as long as we have not mapped them in the light of HE
photons. However, independent of the lateral profile of CRp
acceleration, we expect that the emission profile will be centrally
peaked, as the HE emission results from inelastic collisions of the
CRps with the centrally peaked thermal target material.

The sensitivity of the Whipple 10 m telescopes drops for angular
distances exceeding 0.8 degree from the center of the field of
view. For the Perseus cluster, the temperature map of
\cite{churazov.2003} shows a high-temperature region at about 0.25
degree from the cluster center. As the high-temperature region might
be associated with shocks, this region might emit VHE emission. Our
search for VHE emission does cover this region with high
sensitivity. However, we did not perfom a specialized search for
merger related emission.

In the case of the more distant cluster A2029, our search
region of 0.8 degree radius covers a physical region of 4.2 Mpc
radius. Thus, our search includes all the cluster emission,
independent of where in the cluster it originates.

The rest of the paper is organized as follows: we describe the Whipple
10 m telescope, the observations, data cleaning procedures, and the
data analysis methods in Section \ref{data}.  The main results of this
study are a search for point source emission from localized sources in
the clusters and a search for diffuse emission from the ICM.  We
present these results in Section \ref{results}, and discuss them in
Section \ref{discussion}.  Reported uncertainties are one standard
deviation and upper limits are given at the 90\% confidence level,
unless otherwise stated.  In the rest of the paper, we assume Hubble's
constant $H_0=70$~km~sec$^{-1}$~Mpc$^{-1}$, the dark energy density
$\Omega_\Lambda=0.7$ and the total matter density $\Omega_{\rm
M}=0.3$.

\section{\label {data} Data and Analysis}

\subsection{Instrumentation and Data Sets}

TeV observations were taken on clear moonless nights with the Whipple
10 m Cherenkov telescope located on Mount Hopkins, Arizona at an
altitude of 2300 meters above sea level.  This telescope detects high
energy photons by imaging the flashes of Cherenkov light emitted by
secondary particles in gamma ray induced air showers.  The Whipple 10
m segmented mirror focuses the faint UV/blue Cherenkov flashes onto
a camera consisting of photomultiplier tube pixels.  The Whipple
telescope, including the current camera, have been described elsewhere
\citep{finley.2001}.

\begin{figure}
\plotone{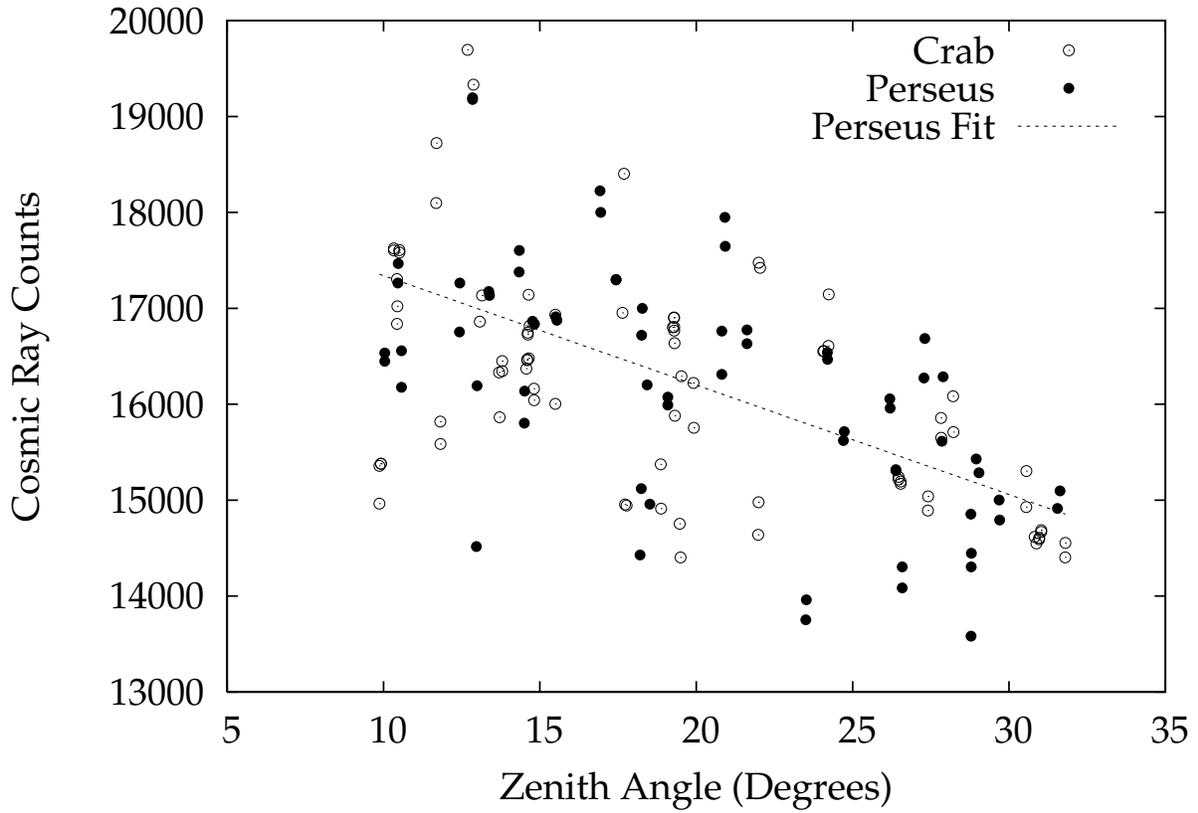}
\caption{Cosmic Ray counts on a run by run basis versus zenith angle.
  Shown are data from the Crab Nebula (open circles) and the Perseus
  Cluster (closed circles).  We fitted each group of data (see the
  Perseus Fit line for an example) to show the dependence of the rate
  on the zenith angle and rejected any runs that deviated by more than
  10\%.\label{cleaning} }
\end{figure}

We observed the Perseus cluster between August 16, 2004 and February
5, 2005 (UT).  Data were taken as pairs of 28 minute runs. An ON run
pointed at the source was followed by an OFF run at the same azimuth
and elevation but offset 7.50$^\circ$ (30 minutes) in right ascension
for background subtraction.  Removing runs with low raw rates
(indicative of poor sky conditions) and mismatched ON/OFF pairs
(indicative of differing sky conditions between ON and OFF runs)
resulted in 29 ON/OFF pairs for analysis.  The cluster Abell 2029 was
observed between March 7, 2003 and May 5, 2003 (UT) resulting in 14
ON/OFF pairs. A number of observations of the Crab Nebula (a
``Standard Candle'' in TeV gamma-ray astronomy) were taken to
determine the detection efficiency and angular resolution for various
points on the camera.  Figure \ref{cleaning} illustrates the cosmic
ray rates of each run versus the zenith angle for both the Perseus and
Crab observations. In this analysis, we only use those runs that
deviate by less than 10\% from the expected rate. Table
\ref{tab:datasets} details the duration and observing season of the
various data sets.

\subsection{Standard Analysis}

The data were analyzed using the standard
  2$^{\rm{nd}}$-moment-parameterization technique \citep{hillas.1985}.
  We identify gamma-ray events and suppress background cosmic ray
  events by applying gamma-ray selection criteria
  (EZCuts2004,~see~\cite{kosack.2005}), designed to be independent of
  zenith angle and energy and well suited for the analysis of extended
  sources. The 2D arrival direction of each gamma-ray event was
  calculated from the orientation and elongation of the Cherenkov
  light distribution in the camera \citep{buckley.1998}. We estimate
  that the mean energy threshold for the Whipple 10 m to be
  approximately 400~GeV \citep{finley.2001}. More detailed
  descriptions of Whipple observing modes and analysis procedures have
  been given by \cite{weekes.1996}, \cite{punch.1991}, and
  \cite{reynolds.1993}.

\subsection{Cluster Specific Analysis}

In this section we describe the specific analysis techniques applied
to the clusters, including the method used to search for point sources
within each cluster. Based upon the expected lateral emission
profiles, we then discuss the examination of the cluster for diffuse
emission.

\begin{figure}
\plotone{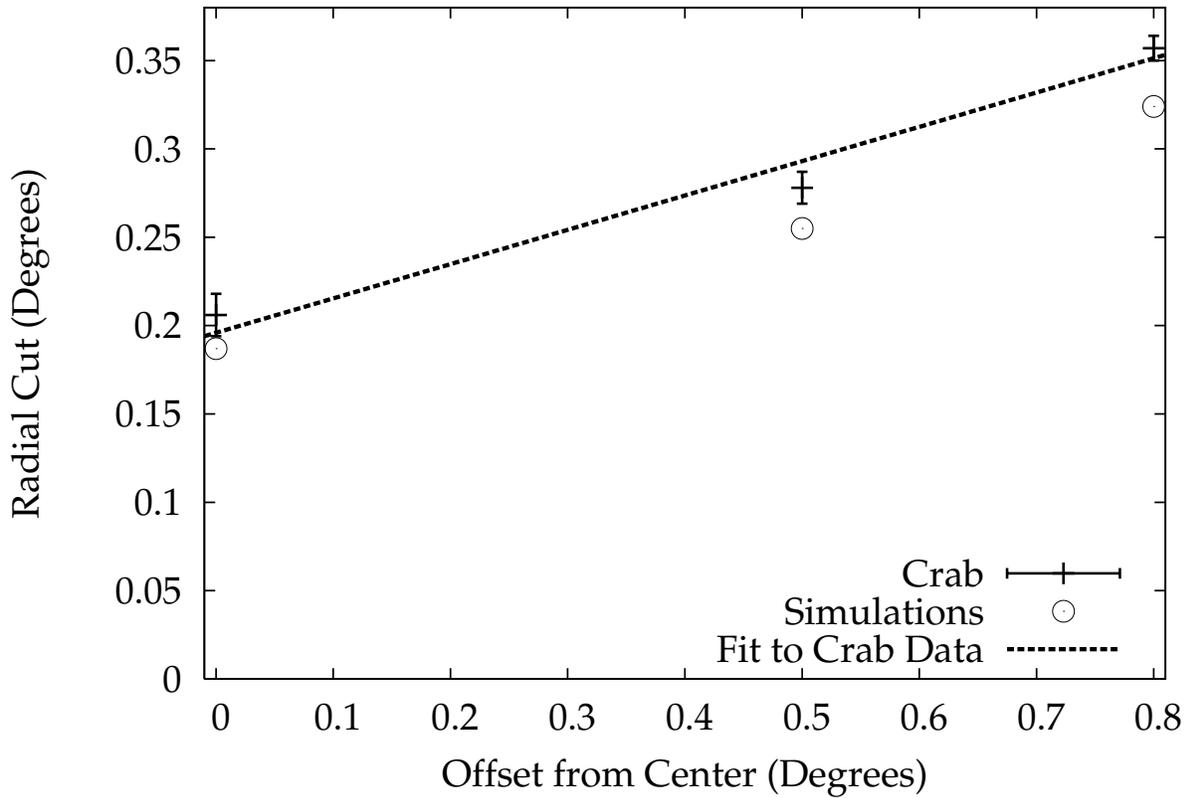}
\caption{Optimal angular cut for different source locations on the
  camera. All the cuts accept $\sim$ 50\% of all the triggered Crab
  events. Shown are the results from the Crab observations in 2004
  (crosses) and from Monte Carlo simulations (circles). The cut
  increases further from the center due to the loss in resolution.
  The fit to the Crab data (dashed line) was used to search for point
  sources in the field-of-view. \label{cut} }
\end{figure}

\begin{figure}
\plotone{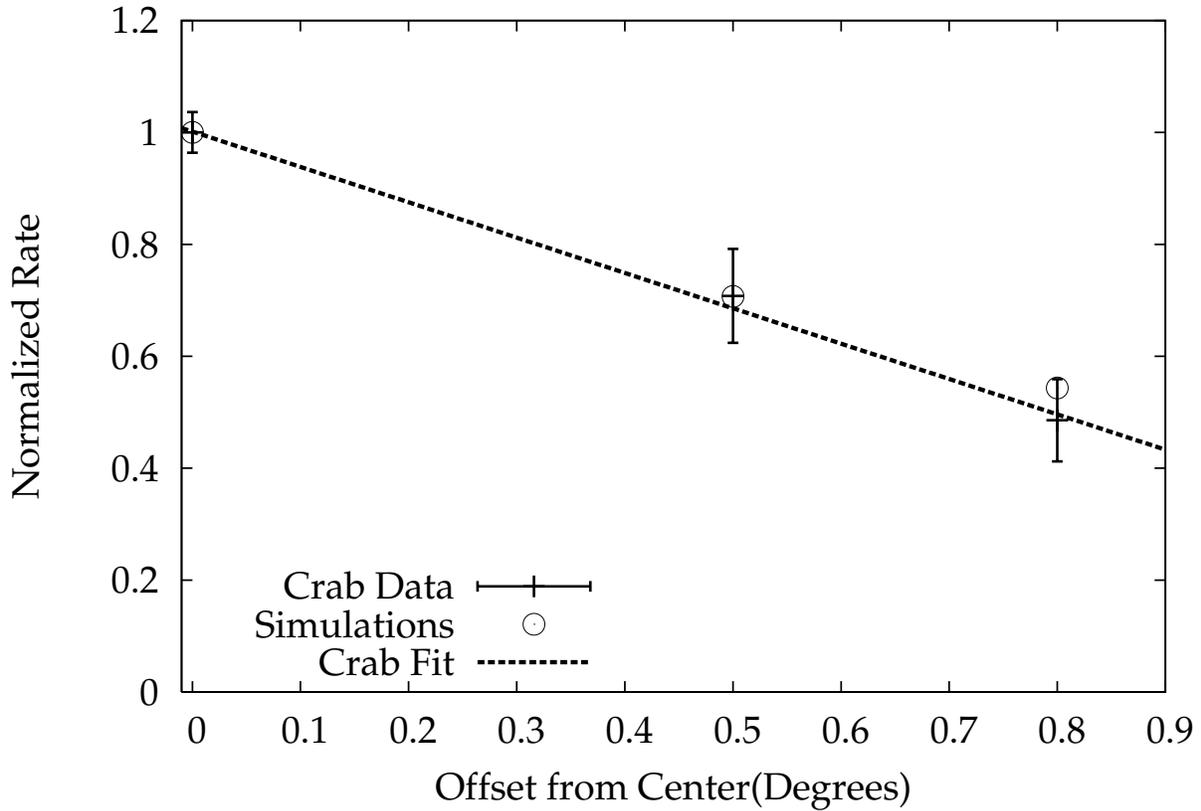}
\caption{Crab gamma-ray rate normalized to 1.0 at zero offset
  (crosses) versus offset from the center of the camera using the
  optimized cut found in Figure \ref{cut}.  The fit to these data
  (dashed line) was used to calculate the upper limit for point
  sources within the field-of-view.  Also shown are the results from
  Monte Carlo simulations (circles) that match the observational data
  very well. At the center of the field-of-view, the detection rate is
  1.9 events per minute.\label{rate} }
\end{figure}

In order to search for point sources within the field-of-view, the
resolution and detection efficiency need to be known to good accuracy
at all locations on the camera. Every search for extended emission
should be preceded by a search for point sources. If there are point
sources, the corresponding sky regions should be excluded from the
search for extended emission.  We used an empirical method based upon
data from the Crab Nebula that were taken during the same months as
the Perseus and Abell 2029 data. The background-subtracted Crab data
were binned by the square of the distance of the reconstructed shower
direction from the location of the Crab Nebula (so as to eliminate any
solid angle dependence) and fitted with an exponential.  These fits
gave us a direct measurement of the resolution of the camera for a
point source at different locations within the field-of-view. From
these same data we determined an optimal angular cut based on the
integral number of excess and background counts as a function of
angular distance from the source location. By calculating the
gamma-ray rate at the different offsets, we also determined how the
efficiency of the camera falls off towards the edges.  This empirical
method was compared to Monte Carlo simulations of centered and offset
data.  The Monte Carlo
code\footnote{http://www.physics.utah.edu/gammaray/GrISU/} simulates
atmospheric Cherenkov showers and calculates the response of the
Whipple detector.  The simulated data have the same format as the
experimental data and were analyzed using the same methods as those
applied to the real data.  We produced a simulated shower set with a
differential spectral index of -2.5 and fed this through the detector
simulations for different source offsets and compared this with
observations.  Figure \ref{cut} shows the optimal angular cut at the
three different offsets.  The optimal cut was used to determine the
total number of events originating from a specific point in the
field-of-view. This cut increases from 0.2$^\circ$ at the camera
center to 0.35$^\circ$ at a 0.8$^\circ$ source offset due to the
poorer angular resolution towards the camera edge.  Figure \ref{rate}
shows the normalized gamma-ray rate for the source located at the
various offsets using the cut from Figure \ref{cut}.  Compared to the
center of the field-of-view, the rate decreases by 40\% at 0.8$^\circ$
from the center due to the loss in detection efficiency.  The
simulated data rates and optimized cuts agree well with the
experimental results. Since the efficiency of the detector falls off
above a radius of 0.8$^\circ$, we only use the central 1.6$^\circ$
diameter region.  If TeV emission mimics the thermal surface
brightness we would see almost all of the emission expected.
Unfortunately, our search has only very limited sensitivity beyond the
central 0.8$^\circ$ from the center of the field of view.

We then searched over the central 0.8$^\circ$ radius region of the
field-of-view of the camera for point sources within the clusters.  At
every point in the field-of-view, we applied the optimal cut as
specified and calculated the excess or deficit of candidate gamma rays
from the data.  We normalized the excess or deficit counts to the
experimentally measured Crab rates from the same observing season. We
then used this flux and its error to calculate a Bayesian upper limit
on the flux \citep{helene.1983}, taking into account the statistical
error for the Crab event rate.

\begin{figure}
\plotone{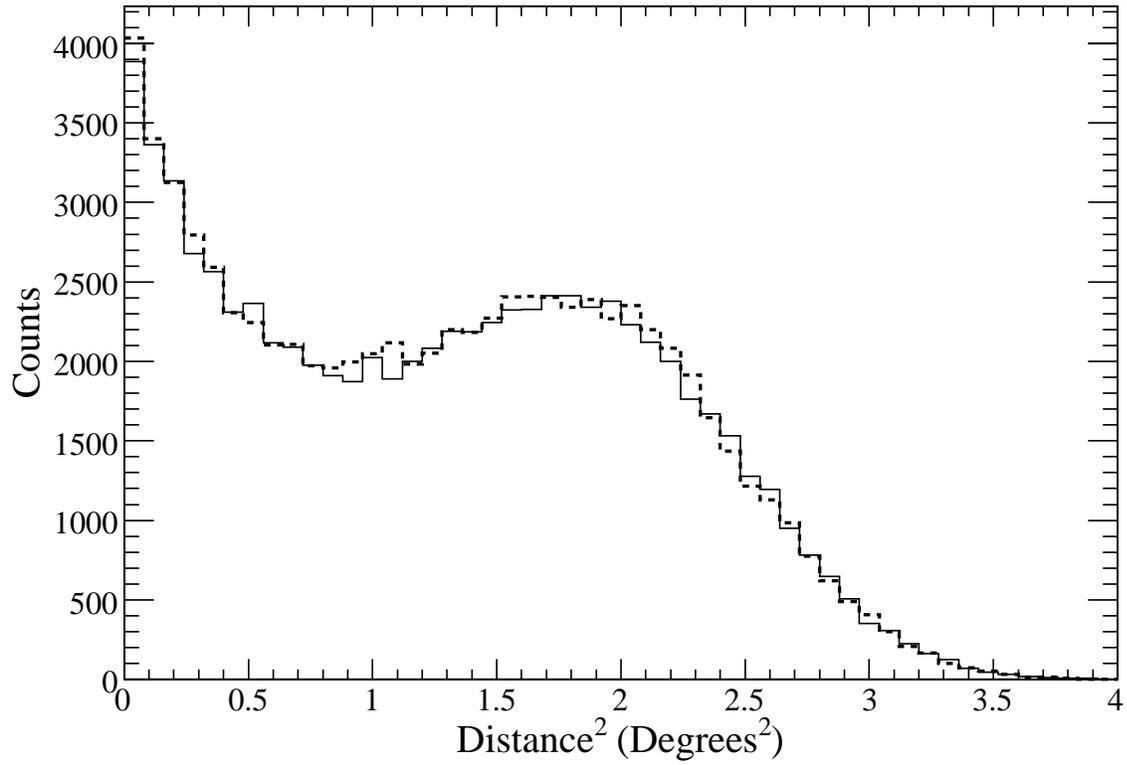}
\caption{Number of Whipple 10~m Perseus observation events versus the
  distance of the estimated arrival direction from the center of the
  field-of-view squared.  The dashed line shows the OFF counts and the
  solid line the ON counts.  There is a good match between the ON and
  the OFF data out to the edge of the camera, and no excess from the
  cluster is detected.\label {theta} }
\end{figure}

To search for extended emission from the Perseus cluster, we assumed
that the TeV gamma-ray surface brightness mimics that of the thermal
X-ray emission seen by Chandra \citep{sanders.2005} and BeppoSAX
\citep{nevalainen.2004} which arises from interactions of the thermal
protons in the cluster.  The X-ray surface brightness can be modeled
as a double-$\beta$ profile:
\begin{equation}
\label{beta}
\Sigma(r) \propto \left(\sum_{i=1}^2 a_i(1+\frac{r^2}{r^2_i})^{-3\beta_i/2}\right)^2
\end{equation}
where $\Sigma(r)$ is the surface brightness and $a_i, r_i$ and
$\beta_i$ are isobaric model parameters \citep{pfrommer.2004}.  The
values of these parameters, based on results from \cite{churazov.2003}
and \cite{struble.1999}, can be found in Table \ref{tab:king}. The
emission will continue out to the accretion shock which is expected to
occur at $\sim$2.2$^\circ$ from the cluster center.  Assuming the
double-$\beta$ profile, we estimate that our angular cut of
0.3$^\circ$ from the cluster center optimizes the sensitivity of the
search for cluster emission.  A fraction of 95\% of the total cluster
emission comes from within 0.3$^\circ$ from the cluster.  Figure
\ref{theta} shows the ON and OFF data after analysis and cleaning
plotted versus the distance from the center of the field-of-view
squared. There is an excellent match between the ON and OFF data and
no obvious excess out to the edge of the field-of-view.

The X-ray surface brightness is better modeled in the case of Abell
2029 by a single-$\beta$ King profile \citep{king.1972} given by
\begin{equation}
\label{king}
\Sigma(r) \propto a_i(1+\frac{r^2}{r^2_1})^{-3\beta_1 + 1/2}.
\end{equation}
The model parameters are found in Table \ref{tab:king} and are from
\cite{jones.1984} based upon observations made with the
\textit{Einstein} observatory.  We chose the \textit{Einstein}
observations over more recent observations by \textit{Chandra} due to
the larger field-of-view of \textit{Einstein}. For this cluster, the
X-ray emission continues out to $\sim$1.0$^\circ$ from the center of
the cluster, and 96\% of the emission comes from the central
0.3$^\circ$.

\begin{figure}
\epsscale{.75}
\plotone{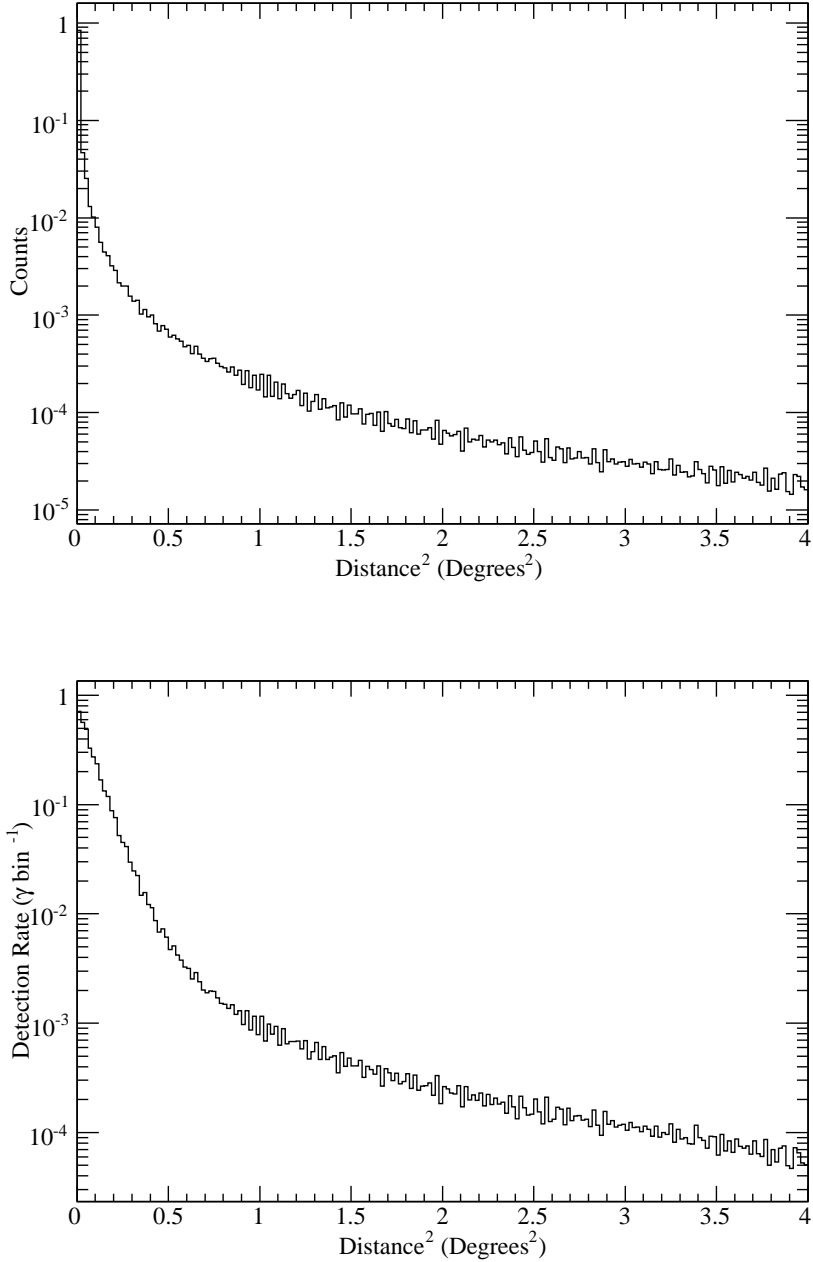} 
\caption{The Upper figure shows the expected count distribution for
  the Whipple telescope based upon the double-$\beta$ profile for the
  Perseus cluster (Equation \ref{beta}) normalized so that the area
  under the curve is 1.0.  The lower figure is this count distribution
  convolved with the angular resolution and the Crab detection rate of
  the Whipple 10 m telescope. The lower plot can be integrated to give
  the total expected signal from the Perseus cluster if it shines with
  the flux of the Crab Nebula. A fact to note is that almost all the
  expected emission arises from within 0.3$^\circ$ of the cluster
  core.\label{profile}}
\end{figure}

We derived quantitative upper limits by normalizing these profiles
over the field-of-view of the camera.  We then convolved the expected
emission by the point spread function of the Whipple telescope,
multiplied by the offset-dependent Crab detection rate. The method
generates a map of the expected detection rate, assuming that the
entire cluster produces the same TeV flux as the Crab Nebula. Figure
\ref{profile} shows the expected emission based on the double-$\beta$
profile for the Perseus cluster at various stages in the analysis
process.  The rate map and actual excess were integrated over the
inner 0.3$^\circ$ and these two values were used to determine the
upper limit on the diffuse TeV flux from the entire cluster in units
of the Crab flux.  We also computed upper limits by integrating counts
over the most sensitive 0.8$^\circ$ region of the camera.

\section{\label {results} Results}
\begin{figure}
\epsscale{1}
\plotone{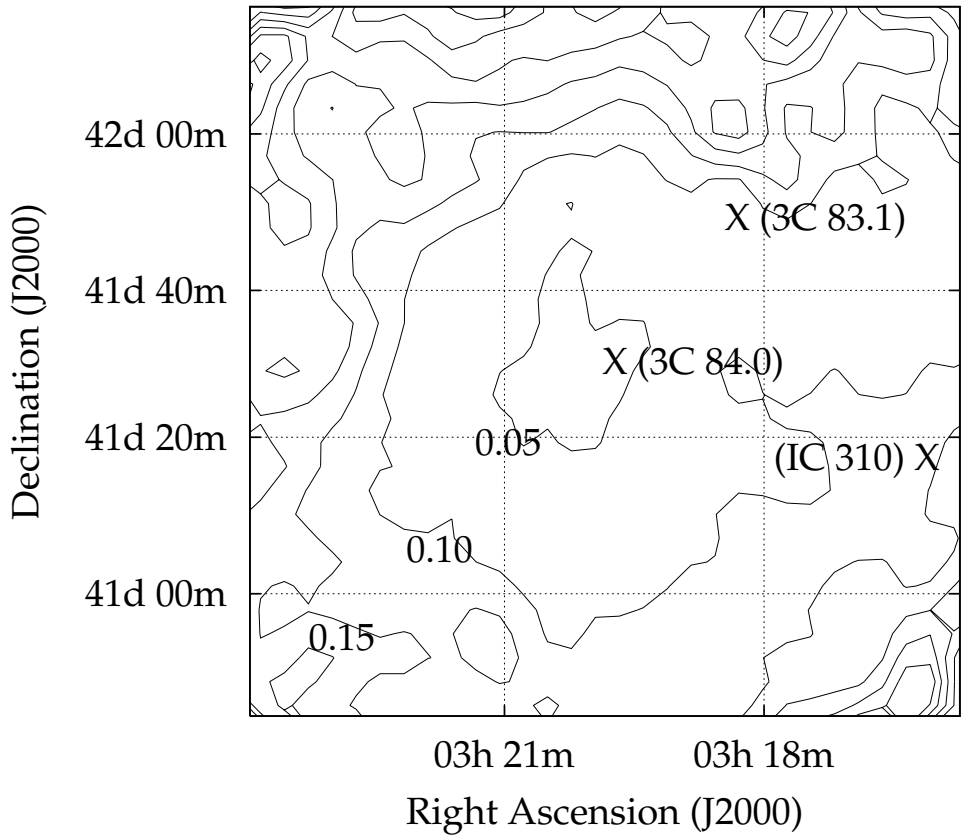}
\caption{Gamma-ray flux upper limit map (90\% confidence level) from
  point sources of the inner 1 degree of the Perseus cluster of
  galaxies.  The scale is in units of the flux from Crab Nebula with
  each contour step equal to 0.05 times the Crab flux.  The
  approximate location of the radio sources found in Table~\ref{radio}
  are shown.\label {upper} }
\end{figure}

For the Perseus cluster, Figure \ref{theta} shows that there is no
excess detected in the field-of-view of the camera. Using a radial cut
of 0.3$^\circ$, our analysis results in a significance of -2.1
standard deviations and an upper limit on the diffuse emission of 13\%
of the Crab flux ($7.4\times10^{-12}$ ergs~cm$^{-2}$~s$^{-1}$). Figure
\ref{upper} is a map of upper limits on the point source emission. All
of the upper limits are below 0.45 Crab, and most (80\%) are below
0.05 Crab. Table \ref{radio} lists the upper limits at the locations
of the three radio galaxies associated with spectroscopically
identified cluster galaxies.  The upper limit on the TeV emission from
the central galaxy, NGC 1275, is 4.0\% of the Crab flux
($2.7\times10^{-12}$ ergs~cm$^{-2}$~s$^{-1}$).

\begin{figure}
\plotone{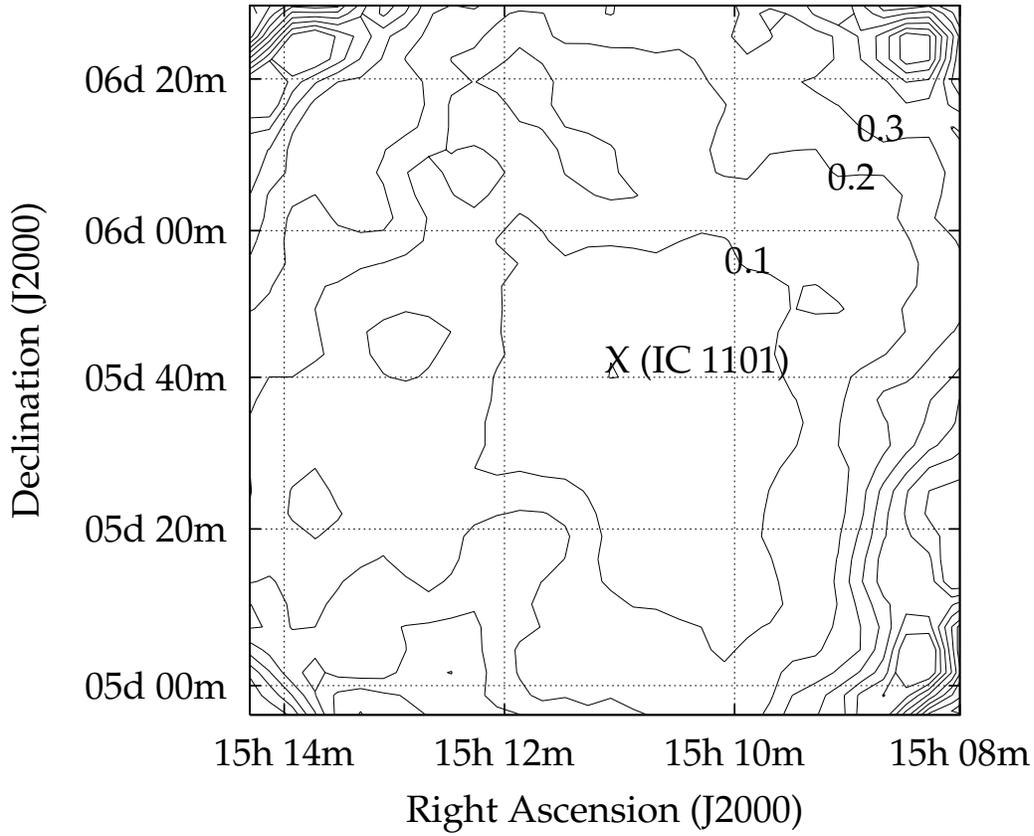}
\caption{Gamma-ray flux upper limit map (90\% confidence level) from
  point sources of the inner 1 degree of the Abell 2029 cluster of
  galaxies.  The scale is in units of the flux from Crab Nebula with
  each contour step equal to 0.1 times the Crab flux.  Select contours
  are labeled. The location of the central brightest radio galaxy is
  shown.  \label {upper:a2029}}
\end{figure}

Abell 2029 does not show any evidence for point source or extended
emission. Figure \ref{upper:a2029} shows a map of upper limits on the
point source emission. All of the upper limits are below 1.1 Crab with
the majority (80\%) below 0.1 Crab. Table \ref{radio} lists an upper
limit of 13\% of the Crab flux ($14\times10^{-12}$
ergs~cm$^{-2}$~s$^{-1}$) for the central radio galaxy. Within
0.3$^\circ$ from the camera center, we find a deficit of 13 counts
with a statistical significance of -0.15 standard deviations. Assuming
the emission profile of Abell 2029 follows Eq. (\ref{king}), we
compute an upper limit on the diffuse emission of 14\% of the Crab
flux ($16\times10^{-12}$ ergs~cm$^{-2}$~s$^{-1}$) . Table \ref{limits}
gives a summary of the various upper limits for each cluster.  All
upper limits discussed in this paper have been computed for the
gamma-ray emission from within 0.2$^\circ$, 0.3$^\circ$ and
0.8$^\circ$ angular distance from the cluster core.  Flux upper limits
have been scaled based upon the assumed spectral shape after
absorption.

\section{\label{discussion}Interpretation and Discussion}

\begin{figure}
\plotone{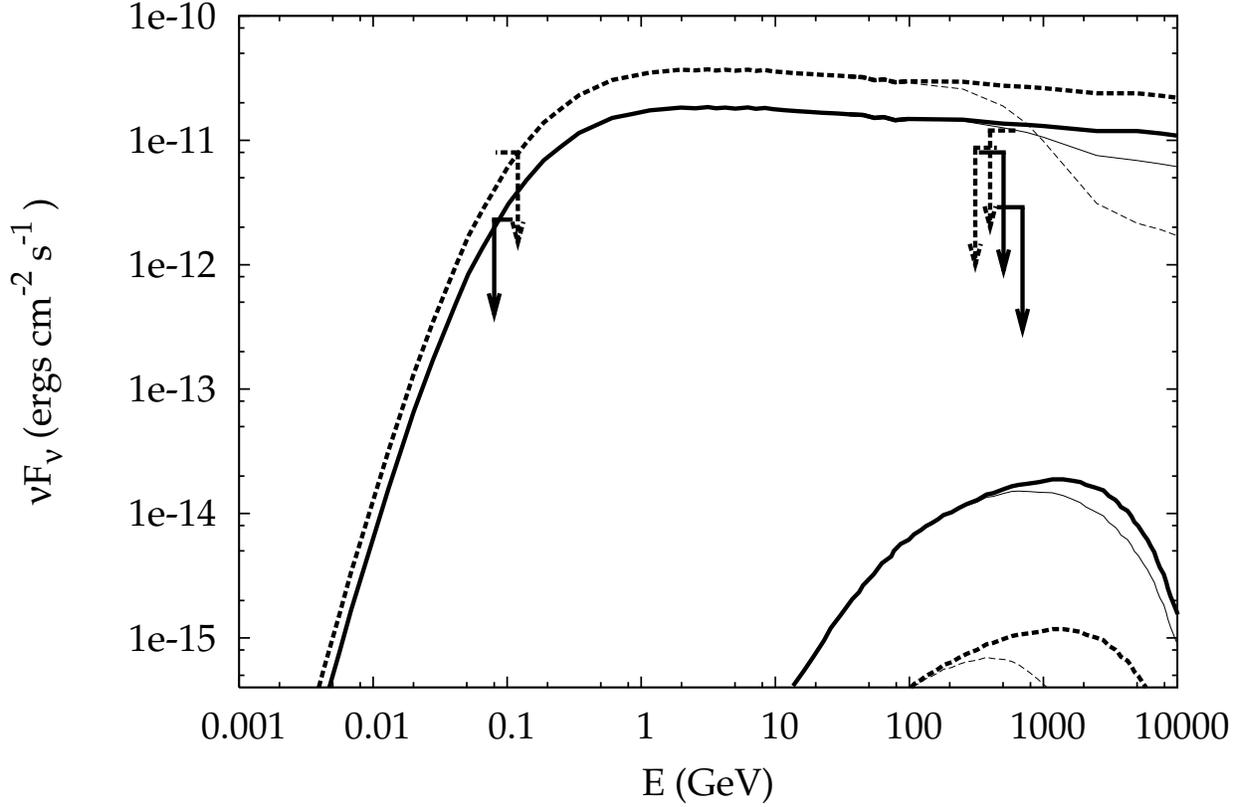}
\caption{In this plot, the solid lines correspond to the Perseus
  cluster and the dashed to Abell 2029. The Whipple 90\% upper limits
  on the emission from the clusters are plotted at 400 GeV (offset to
  improve readability) with the higher values in each case
  corresponding to an angular cut of 0.3$^\circ$ (optimized for the
  search for diffuse CRp emission) and the lower to a cut of
  0.2$^\circ$ (optimized for the search for point sources and dark
  matter).  The upper solid and dashed lines show the CRp induced pion
  decay gamma-ray emission \citep{pfrommer.2004} normalized to the
  {\it EGRET} 100 MeV upper limit (shown at 100MeV). Also plotted (the
  lower flux emission at the bottom right) is the dark matter emission
  derived under the assumption that the TeV gamma-ray signal from the
  galactic center originates from the annihilation of an 18 TeV
  neutralino \citep{horns.2005} which should be compared to the point
  source upper limits (0.2$^\circ$ cut).  The thin lines emanating
  from the pion and dark matter spectra show the effect of
  extragalactic extinction owing to pair production processes.\label
  {flux_predictions}}
\end{figure}

Figure \ref{flux_predictions} shows our upper limits on TeV emission
from the two clusters and compares them to previous upper limits from
{\it EGRET} \citep{reimer.2003}, with the results of model
calculations. The lines show models of the CRp induced gamma-ray
emission normalized to the {\it EGRET} upper limits, assuming a CRp
spectrum with differential spectral index of -2.1
\citep{pfrommer.2004}. This index is a reasonable choice of the source
spectrum because galaxy clusters are not ``leaky'' and retain all CRp,
unlike our Galaxy where leakage of high energy CRp is thought to
steepen the source spectrum of -2.1 to the observed value of -2.7. If
we assume a spectral index of -2.3, the Whipple and {\it EGRET} upper
limits are equivalent. Also shown on this plot (as a thinner extension
to the main lines) is a prediction of the emission modified by
extragalactic extinction owing to pair production processes of TeV
photons with photons of the cosmic infrared background ($\gamma_{TeV}
+ \gamma_{CIB} \rightarrow e^+ + e^-$). The extinction calculation
assumes the phenomenological background model (``P0.45'') of
\cite{aharonian.2005}. Extragalactic extinction has only a minor
impact on the flux predictions for Perseus owing to its low
redshift. However, Abell 2029 is significantly farther away and
extinction does influence the observed spectrum which we take into
consideration when calculating upper limits. The Whipple upper limits
(this publication) lie by factors of 4.6 (Perseus) and 4.2 (Abell
2029) below the model extrapolations. If the CRp spectrum indeed
follows a power law distribution with differential spectral index of
-2.1 up to multi-TeV energies, the calculations of
\cite{pfrommer.2004} together with our results imply that the
non-thermal CRp energy density is less than 7.9\% of the thermal
energy density for the Perseus cluster.

The lower lines in Figure \ref{flux_predictions} show the expected
emission from dark matter annihilations derived under the optimistic
assumption that the TeV emission from the galactic center
\citep{aharonian.2004,kosack.2004,tsuchiya.2004,horns.2005} originates
entirely from such annihilations.  We scale the gamma-ray flux from
the galactic center by computing the expected annihilation signal for
the Galactic Center, the Perseus cluster and Abell 2029 from a
Navarro, Frenk and White (NFW) halo \citep{bergstrom.1998} with $\rho
\propto (r/r_s(1+r/r_s)^2)^{-1}$, virial radius $r_s \simeq 290$~kpc,
a halo mass of $4\times10^{14}$, a distance of 75 Mpc and an NFW
concentration parameter of $c \simeq 4$. We find that the best
sensitivity (signal to background noise ratio) is obtained if we use
the same radial cut, $\theta = 0.3^\circ$, as for the search for point
sources (reducing background from misidentified CR air showers). The
expected dark matter signal lies two orders of magnitude or more below
our upper limits.  We will not see dark matter emission even if all of
the TeV galactic center emission is dark matter in origin.  Thus, we
do not provide any new constraints on TeV galactic center emission.
Secondly, our calculations show that the most promising region to
observe dark matter is the galactic center.

Though we did not detect significant TeV gamma rays from these two
clusters of galaxies, we are able to determine two different types of
upper limits on the emission: from point sources within the cluster
and upper limits on the extended emission. Long duration observations
with the more sensitive TeV telescopes VERITAS, HESS, MAGIC and
CANGAROO III and the GeV telescope GLAST will be critical for
determining whether cluster are emitters of high energy gamma rays.

\textit{Acknowledgments:} This research is supported by grants from
the U.S. Department of Energy, the U.S. National Science Foundation,
the Smithsonian Institution, by NSERC in Canada, by Science Foundation
Ireland, and by PPARC in the UK.  H.K. acknowledges the support of the
DOE in the framework of the Outstanding Junior Investigator Award.
J.S.P. acknowledges the support by the Dean of the Arts and Sciences
Graduate School of Washington University in St. Louis through a
dissertation fellowship.  We would also like to thank Jonathan Katz
for his comments and suggestions.

\bibliography{ms.bib}

\begin{thebibliography}{48}
\expandafter\ifx\csname natexlab\endcsname\relax\def\natexlab#1{#1}\fi

\bibitem[{{Aharonian} {et~al.}(2004){Aharonian}, {Akhperjanian}, {Aye},
  {Bazer-Bachi}, {Beilicke}, {Benbow}, {Berge}, {Berghaus}, {Bernl{\" o}hr},
  {Bolz}, {Boisson}, {Borgmeier}, {Breitling}, {Brown}, {Bussons Gordo},
  {Chadwick}, {Chitnis}, {Chounet}, {Cornils}, {Costamante}, {Degrange},
  {Djannati-Ata{\" i}}, {O'C.~Drury}, {Ergin}, {Espigat}, {Feinstein},
  {Fleury}, {Fontaine}, {Funk}, {Gallant}, {Giebels}, {Gillessen}, {Goret},
  {Guy}, {Hadjichristidis}, {Hauser}, {Heinzelmann}, {Henri}, {Hermann},
  {Hinton}, {Hofmann}, {Holleran}, {Horns}, {de Jager}, {Jung}, {Kh{\' e}lifi},
  {Komin}, {Konopelko}, {Latham}, {Le Gallou}, {Lemoine}, {Lemi{\` e}re},
  {Leroy}, {Lohse}, {Marcowith}, {Masterson}, {McComb}, {de Naurois}, {Nolan},
  {Noutsos}, {Orford}, {Osborne}, {Ouchrif}, {Panter}, {Pelletier}, {Pita},
  {Pohl}, {P{\" u}hlhofer}, {Punch}, {Raubenheimer}, {Raue}, {Raux}, {Rayner},
  {Redondo}, {Reimer}, {Reimer}, {Ripken}, {Rivoal}, {Rob}, {Rolland},
  {Rowell}, {Sahakian}, {Saug{\' e}}, {Schlenker}, {Schlickeiser}, {Schuster},
  {Schwanke}, {Siewert}, {Sol}, {Steenkamp}, {Stegmann}, {Tavernet}, {Th{\'
  e}oret}, {Tluczykont}, {van der Walt}, {Vasileiadis}, {Vincent}, {Visser},
  {V{\" o}lk}, \& {Wagner}}]{aharonian.2004}
{Aharonian}, F., {Akhperjanian}, A.~G., {Aye}, K.-M., {Bazer-Bachi}, A.~R.,
  {Beilicke}, M., {Benbow}, W., {Berge}, D., {Berghaus}, P., {Bernl{\" o}hr},
  K., {Bolz}, O., {Boisson}, C., {Borgmeier}, C., {Breitling}, F., {Brown},
  A.~M., {Bussons Gordo}, J., {Chadwick}, P.~M., {Chitnis}, V.~R., {Chounet},
  L.-M., {Cornils}, R., {Costamante}, L., {Degrange}, B., {Djannati-Ata{\" i}},
  A., {O'C.~Drury}, L., {Ergin}, T., {Espigat}, P., {Feinstein}, F., {Fleury},
  P., {Fontaine}, G., {Funk}, S., {Gallant}, Y., {Giebels}, B., {Gillessen},
  S., {Goret}, P., {Guy}, J., {Hadjichristidis}, C., {Hauser}, M.,
  {Heinzelmann}, G., {Henri}, G., {Hermann}, G., {Hinton}, J.~A., {Hofmann},
  W., {Holleran}, M., {Horns}, D., {de Jager}, O.~C., {Jung}, I., {Kh{\'
  e}lifi}, B., {Komin}, N., {Konopelko}, A., {Latham}, I.~J., {Le Gallou}, R.,
  {Lemoine}, M., {Lemi{\` e}re}, A., {Leroy}, N., {Lohse}, T., {Marcowith}, A.,
  {Masterson}, C., {McComb}, T.~J.~L., {de Naurois}, M., {Nolan}, S.~J.,
  {Noutsos}, A., {Orford}, K.~J., {Osborne}, J.~L., {Ouchrif}, M., {Panter},
  M., {Pelletier}, G., {Pita}, S., {Pohl}, M., {P{\" u}hlhofer}, G., {Punch},
  M., {Raubenheimer}, B.~C., {Raue}, M., {Raux}, J., {Rayner}, S.~M.,
  {Redondo}, I., {Reimer}, A., {Reimer}, O., {Ripken}, J., {Rivoal}, M., {Rob},
  L., {Rolland}, L., {Rowell}, G., {Sahakian}, V., {Saug{\' e}}, L.,
  {Schlenker}, S., {Schlickeiser}, R., {Schuster}, C., {Schwanke}, U.,
  {Siewert}, M., {Sol}, H., {Steenkamp}, R., {Stegmann}, C., {Tavernet}, J.-P.,
  {Th{\' e}oret}, C.~G., {Tluczykont}, M., {van der Walt}, D.~J.,
  {Vasileiadis}, G., {Vincent}, P., {Visser}, B., {V{\" o}lk}, H.~J., \&
  {Wagner}, S.~J. 2004, \aap, 425, L13

\bibitem[{{Aharonian} {et~al.}(2005){Aharonian}, {Akhperjanian}, {Bazer-Bachi},
  {Beilicke}, {Benbow}, {Berge}, {Bernl{\"o}hr}, {Boisson}, {Bolz}, {Borrel},
  {Braun}, {Breitling}, {Brown}, {Chadwick}, {Chounet}, {Cornils},
  {Costamante}, {Degrange}, {Dickinson}, {Djannati-Ata{\"i}}, {O'C.~Drury},
  {Dubus}, {Emmanoulopoulos}, {Espigat}, {Feinstein}, {Fontaine}, {Fuchs},
  {Funk}, {Gallant}, {Giebels}, {Gillessen}, {Glicenstein}, {Goret},
  {Hadjichristidis}, {Hauser}, {Heinzelmann}, {Henri}, {Hermann}, {Hinton},
  {Hofmann}, {Holleran}, {Horns}, {Jacholkowska}, {de Jager}, {Kh{\'e}lifi},
  {Komin}, {Konopelko}, {Latham}, {Le Gallou}, {Lemi{\`e}re},
  {Lemoine-Goumard}, {Leroy}, {Lohse}, {Martin}, {Martineau-Huynh},
  {Marcowith}, {Masterson}, {McComb}, {de Naurois}, {Nolan}, {Noutsos},
  {Orford}, {Osborne}, {Ouchrif}, {Panter}, {Pelletier}, {Pita},
  {P{\"u}hlhofer}, {Punch}, {Raubenheimer}, {Raue}, {Raux}, {Rayner}, {Reimer},
  {Reimer}, {Ripken}, {Rob}, {Rolland}, {Rowell}, {Sahakian}, {Saug{\'e}},
  {Schlenker}, {Schlickeiser}, {Schuster}, {Schwanke}, {Siewert}, {Sol},
  {Spangler}, {Steenkamp}, {Stegmann}, {Tavernet}, {Terrier}, {Th{\'e}oret},
  {Tluczykont}, {Vasileiadis}, {Venter}, {Vincent}, {V{\"o}lk}, \&
  {Wagner}}]{aharonian.2005}
{Aharonian}, F., {Akhperjanian}, A.~G., {Bazer-Bachi}, A.~R., {Beilicke}, M.,
  {Benbow}, W., {Berge}, D., {Bernl{\"o}hr}, K., {Boisson}, C., {Bolz}, O.,
  {Borrel}, V., {Braun}, I., {Breitling}, F., {Brown}, A.~M., {Chadwick},
  P.~M., {Chounet}, L.-M., {Cornils}, R., {Costamante}, L., {Degrange}, B.,
  {Dickinson}, H.~J., {Djannati-Ata{\"i}}, A., {O'C.~Drury}, L., {Dubus}, G.,
  {Emmanoulopoulos}, D., {Espigat}, P., {Feinstein}, F., {Fontaine}, G.,
  {Fuchs}, Y., {Funk}, S., {Gallant}, Y.~A., {Giebels}, B., {Gillessen}, S.,
  {Glicenstein}, J.~F., {Goret}, P., {Hadjichristidis}, C., {Hauser}, M.,
  {Heinzelmann}, G., {Henri}, G., {Hermann}, G., {Hinton}, J.~A., {Hofmann},
  W., {Holleran}, M., {Horns}, D., {Jacholkowska}, A., {de Jager}, O.~C.,
  {Kh{\'e}lifi}, B., {Komin}, N., {Konopelko}, A., {Latham}, I.~J., {Le
  Gallou}, R., {Lemi{\`e}re}, A., {Lemoine-Goumard}, M., {Leroy}, N., {Lohse},
  T., {Martin}, J.~M., {Martineau-Huynh}, O., {Marcowith}, A., {Masterson}, C.,
  {McComb}, T.~J.~L., {de Naurois}, M., {Nolan}, S.~J., {Noutsos}, A.,
  {Orford}, K.~J., {Osborne}, J.~L., {Ouchrif}, M., {Panter}, M., {Pelletier},
  G., {Pita}, S., {P{\"u}hlhofer}, G., {Punch}, M., {Raubenheimer}, B.~C.,
  {Raue}, M., {Raux}, J., {Rayner}, S.~M., {Reimer}, A., {Reimer}, O.,
  {Ripken}, J., {Rob}, L., {Rolland}, L., {Rowell}, G., {Sahakian}, V.,
  {Saug{\'e}}, L., {Schlenker}, S., {Schlickeiser}, R., {Schuster}, C.,
  {Schwanke}, U., {Siewert}, M., {Sol}, H., {Spangler}, D., {Steenkamp}, R.,
  {Stegmann}, C., {Tavernet}, J.-P., {Terrier}, R., {Th{\'e}oret}, C.~G.,
  {Tluczykont}, M., {Vasileiadis}, G., {Venter}, C., {Vincent}, P., {V{\"o}lk},
  H.~J., \& {Wagner}, S.~J. 2005, astro-ph/0508073

\bibitem[{{Bergstr{\" o}m} {et~al.}(1998){Bergstr{\" o}m}, {Ullio}, \&
  {Buckley}}]{bergstrom.1998}
{Bergstr{\" o}m}, L., {Ullio}, P., \& {Buckley}, J.~H. 1998, Astroparticle
  Physics, 9, 137

\bibitem[{{Blasi} \& {Colafrancesco}(1999)}]{blasi.1999}
{Blasi}, P. \& {Colafrancesco}, S. 1999, Astroparticle Physics, 12, 169

\bibitem[{{Bowyer} \& {Bergh{\" o}fer}(1998)}]{bowyer.1998}
{Bowyer}, S. \& {Bergh{\" o}fer}, T.~W. 1998, \apj, 506, 502

\bibitem[{{Buckley} {et~al.}(1998){Buckley}, {Akerlof}, {Carter-Lewis},
  {Catanese}, {Cawley}, {Connaughton}, {Fegan}, {Finley}, {Gaidos}, {Hillas},
  {Krennrich}, {Lamb}, {Lessard}, {McEnery}, {Mohanty}, {Quinn}, {Rodgers},
  {Rose}, {Rovero}, {Schubnell}, {Sembroski}, {Srinivasan}, {Weekes}, \&
  {Zweerink}}]{buckley.1998}
{Buckley}, J.~H., {Akerlof}, C.~W., {Carter-Lewis}, D.~A., {Catanese}, M.,
  {Cawley}, M.~F., {Connaughton}, V., {Fegan}, D.~J., {Finley}, J.~P.,
  {Gaidos}, J.~A., {Hillas}, A.~M., {Krennrich}, F., {Lamb}, R.~C., {Lessard},
  R.~W., {McEnery}, J.~E., {Mohanty}, G., {Quinn}, J., {Rodgers}, A.~J.,
  {Rose}, H.~J., {Rovero}, A.~C., {Schubnell}, M.~S., {Sembroski}, G.,
  {Srinivasan}, R., {Weekes}, T.~C., \& {Zweerink}, J. 1998, \aap, 329, 639

\bibitem[{{Churazov} {et~al.}(2003){Churazov}, {Forman}, {Jones}, \& {B{\"
  o}hringer}}]{churazov.2003}
{Churazov}, E., {Forman}, W., {Jones}, C., \& {B{\" o}hringer}, H. 2003, \apj,
  590, 225

\bibitem[{{Colafrancesco}(2001)}]{colafrancesco.2001}
{Colafrancesco}, S. 2001, in AIP Conf. Proc. 587: Gamma 2001: Gamma-Ray
  Astrophysics, 427

\bibitem[{{Condon} {et~al.}(1998){Condon}, {Cotton}, {Greisen}, {Yin},
  {Perley}, {Taylor}, \& {Broderick}}]{condon.1998}
{Condon}, J.~J., {Cotton}, W.~D., {Greisen}, E.~W., {Yin}, Q.~F., {Perley},
  R.~A., {Taylor}, G.~B., \& {Broderick}, J.~J. 1998, \aj, 115, 1693

\bibitem[{{Donahue} {et~al.}(2004){Donahue}, {Voit}, \&
  {Cavagnolo}}]{donahue.2004}
{Donahue}, M.~E., {Voit}, G.~M., \& {Cavagnolo}, K. 2004, American Astronomical
  Society Meeting Abstracts, 205

\bibitem[{{En\ss lin} \& {Biermann}(1998)}]{ensslin.1998}
{En\ss lin}, T.~A. \& {Biermann}, P.~L. 1998, \aap, 330, 90

\bibitem[{{Fegan} {et~al.}(2005){Fegan}, {Badran}, {Bond}, {Boyle}, {Bradbury},
  {Buckley}, {Carter-Lewis}, {Catanese}, {Celik}, {Cui}, {Daniel}, {D'Vali},
  {de la Calle Perez}, {Duke}, {Falcone}, {Fegan}, {Finley}, {Fortson},
  {Gaidos}, {Gammell}, {Gibbs}, {Gillanders}, {Grube}, {Hall}, {Hall}, {Hanna},
  {Hillas}, {Holder}, {Horan}, {Jarvis}, {Jordan}, {Kenny}, {Kertzman},
  {Kieda}, {Kildea}, {Knapp}, {Kosack}, {Krawczynski}, {Krennrich}, {Lang}, {Le
  Bohec}, {Lessard}, {Linton}, {Lloyd-Evans}, {Milovanovic}, {McEnery},
  {Moriarty}, {Mukherjee}, {Muller}, {Nagai}, {Nolan}, {Ong}, {Pallassini},
  {Petry}, {Power-Mooney}, {Quinn}, {Quinn}, {Ragan}, {Rebillot}, {Reynolds},
  {Rose}, {Schroedter}, {Sembroski}, {Swordy}, {Syson}, {Vassiliev}, {Wakely},
  {Walker}, {Weekes}, \& {Zweerink}}]{fegan.2005}
{Fegan}, S.~J., {Badran}, H.~M., {Bond}, I.~H., {Boyle}, P.~J., {Bradbury},
  S.~M., {Buckley}, J.~H., {Carter-Lewis}, D.~A., {Catanese}, M., {Celik}, O.,
  {Cui}, W., {Daniel}, M., {D'Vali}, M., {de la Calle Perez}, I., {Duke}, C.,
  {Falcone}, A., {Fegan}, D.~J., {Finley}, J.~P., {Fortson}, L.~F., {Gaidos},
  J.~A., {Gammell}, S., {Gibbs}, K., {Gillanders}, G.~H., {Grube}, J., {Hall},
  J., {Hall}, T.~A., {Hanna}, D., {Hillas}, A.~M., {Holder}, J., {Horan}, D.,
  {Jarvis}, A., {Jordan}, M., {Kenny}, G.~E., {Kertzman}, M., {Kieda}, D.,
  {Kildea}, J., {Knapp}, J., {Kosack}, K., {Krawczynski}, H., {Krennrich}, F.,
  {Lang}, M.~J., {Le Bohec}, S., {Lessard}, R.~W., {Linton}, E., {Lloyd-Evans},
  J., {Milovanovic}, A., {McEnery}, J., {Moriarty}, P., {Mukherjee}, R.,
  {Muller}, D., {Nagai}, T., {Nolan}, S., {Ong}, R.~A., {Pallassini}, R.,
  {Petry}, D., {Power-Mooney}, B., {Quinn}, J., {Quinn}, M., {Ragan}, K.,
  {Rebillot}, P., {Reynolds}, P.~T., {Rose}, H.~J., {Schroedter}, M.,
  {Sembroski}, G.~H., {Swordy}, S.~P., {Syson}, A., {Vassiliev}, V.~V.,
  {Wakely}, S.~P., {Walker}, G., {Weekes}, T.~C., \& {Zweerink}, J. 2005, \apj,
  624, 638

\bibitem[{{Feretti}(2003)}]{feretti.2003}
{Feretti}, L. 2003, in Texas in Tuscany. XXI Symposium on Relativistic
  Astrophysics, 209--220

\bibitem[{{Finley} {et~al.}(2001){Finley}, {Krennrich}, {Badran}, {Bond},
  {Bradbury}, {Buckley}, {Carter-Lewis}, {Catanese}, {Cui}, {Dunlea}, {Das},
  {de la Calle Perez}, {Fegan}, {Fegan}, {Gaidos}, {Gibbs}, {Gillanders},
  {Hall}, {Hillas}, {Holder}, {Horan}, {Jordan}, {Kertzman}, {Kieda}, {Kildea},
  {Knapp}, {Kosack}, {Lang}, {LeBohec}, {McKernan}, {Moriarty}, {M{\"u}ller},
  {Ong}, {Pallassini}, {Petry}, {Quinn}, {Reay}, {Reynolds}, {Rose},
  {Sembroski}, {Sidwell}, {Stanton}, {Swordy}, {Vassiliev}, {Wakely}, \&
  {Weekes}}]{finley.2001}
{Finley}, J.~P., {Krennrich}, F., {Badran}, H.~M., {Bond}, I.~H., {Bradbury},
  S.~M., {Buckley}, J.~H., {Carter-Lewis}, D.~A., {Catanese}, M., {Cui}, W.,
  {Dunlea}, S., {Das}, D., {de la Calle Perez}, I., {Fegan}, D.~J., {Fegan},
  S.~J., {Gaidos}, J.~A., {Gibbs}, K., {Gillanders}, G.~H., {Hall}, T.~A.,
  {Hillas}, A.~M., {Holder}, J., {Horan}, D., {Jordan}, M., {Kertzman}, M.,
  {Kieda}, D., {Kildea}, J., {Knapp}, J., {Kosack}, K., {Lang}, M.~J.,
  {LeBohec}, S., {McKernan}, B., {Moriarty}, P., {M{\"u}ller}, D., {Ong}, R.,
  {Pallassini}, R., {Petry}, D., {Quinn}, J., {Reay}, N.~W., {Reynolds}, P.~T.,
  {Rose}, H.~J., {Sembroski}, G.~H., {Sidwell}, R., {Stanton}, N., {Swordy},
  S.~P., {Vassiliev}, V.~V., {Wakely}, S.~P., \& {Weekes}, T.~C. 2001, in
  International Cosmic Ray Conference, 2827

\bibitem[{{Fusco-Femiano} {et~al.}(1999){Fusco-Femiano}, {dal Fiume},
  {Feretti}, {Giovannini}, {Grandi}, {Matt}, {Molendi}, \&
  {Santangelo}}]{fusco-femiano.1999}
{Fusco-Femiano}, R., {dal Fiume}, D., {Feretti}, L., {Giovannini}, G.,
  {Grandi}, P., {Matt}, G., {Molendi}, S., \& {Santangelo}, A. 1999, \apjl,
  513, L21

\bibitem[{{Fusco-Femiano} {et~al.}(2004){Fusco-Femiano}, {Orlandini},
  {Brunetti}, {Feretti}, {Giovannini}, {Grandi}, \&
  {Setti}}]{fusco-femiano.2004}
{Fusco-Femiano}, R., {Orlandini}, M., {Brunetti}, G., {Feretti}, L.,
  {Giovannini}, G., {Grandi}, P., \& {Setti}, G. 2004, \apjl, 602, L73

\bibitem[{{Giovannini} \& {Feretti}(2000)}]{giovannini.2000}
{Giovannini}, G. \& {Feretti}, L. 2000, New Astronomy, 5, 335

\bibitem[{{Giovannini} {et~al.}(1993){Giovannini}, {Feretti}, {Venturi}, {Kim},
  \& {Kronberg}}]{giovannini.1993}
{Giovannini}, G., {Feretti}, L., {Venturi}, T., {Kim}, K.-T., \& {Kronberg},
  P.~P. 1993, \apj, 406, 399

\bibitem[{{Giovannini} {et~al.}(1999){Giovannini}, {Tordi}, \&
  {Feretti}}]{giovannini.1999}
{Giovannini}, G., {Tordi}, M., \& {Feretti}, L. 1999, New Astronomy, 4, 141

\bibitem[{{Girardi} {et~al.}(1998){Girardi}, {Giuricin}, {Mardirossian},
  {Mezzetti}, \& {Boschin}}]{girardi.1998}
{Girardi}, M., {Giuricin}, G., {Mardirossian}, F., {Mezzetti}, M., \&
  {Boschin}, W. 1998, \apj, 505, 74

\bibitem[{{Helene}(1983)}]{helene.1983}
{Helene}, O. 1983, Nuclear Instruments and Methods in Physics Research, 212,
  319

\bibitem[{{Hillas}(1985)}]{hillas.1985}
{Hillas}, A.~M. 1985, in International Cosmic Ray Conference, 445--448

\bibitem[{{Horns}(2005)}]{horns.2005}
{Horns}, D. 2005, Physics Letters B, 607, 225

\bibitem[{{Jones} \& {Forman}(1984)}]{jones.1984}
{Jones}, C. \& {Forman}, W. 1984, \apj, 276, 38

\bibitem[{{Kang} \& {Jones}(2005)}]{kang.2005}
{Kang}, H. \& {Jones}, T.~W. 2005, \apj, 620, 44

\bibitem[{{Kawasaki} \& {Totani}(2002)}]{kawasaki.2002}
{Kawasaki}, W. \& {Totani}, T. 2002, \apj, 576, 679

\bibitem[{{Kempner} \& {Sarazin}(2001)}]{kempner.2001}
{Kempner}, J.~C. \& {Sarazin}, C.~L. 2001, \apj, 548, 639

\bibitem[{{King}(1972)}]{king.1972}
{King}, I.~R. 1972, \apjl, 174, L123

\bibitem[{{Kosack}(2005)}]{kosack.2005}
{Kosack}, K. 2005, PhD thesis, Washington University in St. Louis

\bibitem[{{Kosack} {et~al.}(2004){Kosack}, {Badran}, {Bond}, {Boyle},
  {Bradbury}, {Buckley}, {Carter-Lewis}, {Celik}, {Connaughton}, {Cui},
  {Daniel}, {D'Vali}, {de la Calle Perez}, {Duke}, {Falcone}, {Fegan}, {Fegan},
  {Finley}, {Fortson}, {Gaidos}, {Gammell}, {Gibbs}, {Gillanders}, {Grube},
  {Gutierrez}, {Hall}, {Hall}, {Hanna}, {Hillas}, {Holder}, {Horan}, {Jarvis},
  {Jordan}, {Kenny}, {Kertzman}, {Kieda}, {Kildea}, {Knapp}, {Krawczynski},
  {Krennrich}, {Lang}, {Le Bohec}, {Linton}, {Lloyd-Evans}, {Milovanovic},
  {McEnery}, {Moriarty}, {Muller}, {Nagai}, {Nolan}, {Ong}, {Pallassini},
  {Petry}, {Power-Mooney}, {Quinn}, {Quinn}, {Ragan}, {Rebillot}, {Reynolds},
  {Rose}, {Schroedter}, {Sembroski}, {Swordy}, {Syson}, {Vassiliev}, {Wakely},
  {Walker}, {Weekes}, \& {Zweerink}}]{kosack.2004}
{Kosack}, K., {Badran}, H.~M., {Bond}, I.~H., {Boyle}, P.~J., {Bradbury},
  S.~M., {Buckley}, J.~H., {Carter-Lewis}, D.~A., {Celik}, O., {Connaughton},
  V., {Cui}, W., {Daniel}, M., {D'Vali}, M., {de la Calle Perez}, I., {Duke},
  C., {Falcone}, A., {Fegan}, D.~J., {Fegan}, S.~J., {Finley}, J.~P.,
  {Fortson}, L.~F., {Gaidos}, J.~A., {Gammell}, S., {Gibbs}, K., {Gillanders},
  G.~H., {Grube}, J., {Gutierrez}, K., {Hall}, J., {Hall}, T.~A., {Hanna}, D.,
  {Hillas}, A.~M., {Holder}, J., {Horan}, D., {Jarvis}, A., {Jordan}, M.,
  {Kenny}, G.~E., {Kertzman}, M., {Kieda}, D., {Kildea}, J., {Knapp}, J.,
  {Krawczynski}, H., {Krennrich}, F., {Lang}, M.~J., {Le Bohec}, S., {Linton},
  E., {Lloyd-Evans}, J., {Milovanovic}, A., {McEnery}, J., {Moriarty}, P.,
  {Muller}, D., {Nagai}, T., {Nolan}, S., {Ong}, R.~A., {Pallassini}, R.,
  {Petry}, D., {Power-Mooney}, B., {Quinn}, J., {Quinn}, M., {Ragan}, K.,
  {Rebillot}, P., {Reynolds}, P.~T., {Rose}, H.~J., {Schroedter}, M.,
  {Sembroski}, G.~H., {Swordy}, S.~P., {Syson}, A., {Vassiliev}, V.~V.,
  {Wakely}, S.~P., {Walker}, G., {Weekes}, T.~C., \& {Zweerink}, J. 2004,
  \apjl, 608, L97

\bibitem[{{Krawczynski}(2002)}]{krawczynski.2002}
{Krawczynski}, H. 2002, \apjl, 569, L27

\bibitem[{{Lieu} {et~al.}(1999){Lieu}, {Ip}, {Axford}, \&
  {Bonamente}}]{lieu.1999}
{Lieu}, R., {Ip}, W.-H., {Axford}, W.~I., \& {Bonamente}, M. 1999, \apjl, 510,
  L25

\bibitem[{{Lieu} {et~al.}(1996){Lieu}, {Mittaz}, {Bowyer}, {Breen}, {Lockman},
  {Murphy}, \& {Hwang}}]{lieu.1996}
{Lieu}, R., {Mittaz}, J.~P.~D., {Bowyer}, S., {Breen}, J.~O., {Lockman}, F.~J.,
  {Murphy}, E.~M., \& {Hwang}, C.-Y. 1996, Science, 274, 1335

\bibitem[{{Markevitch} {et~al.}(1998){Markevitch}, {Forman}, {Sarazin}, \&
  {Vikhlinin}}]{markevitch.1998}
{Markevitch}, M., {Forman}, W.~R., {Sarazin}, C.~L., \& {Vikhlinin}, A. 1998,
  \apj, 503, 77

\bibitem[{{Nevalainen} {et~al.}(2004){Nevalainen}, {Oosterbroek}, {Bonamente},
  \& {Colafrancesco}}]{nevalainen.2004}
{Nevalainen}, J., {Oosterbroek}, T., {Bonamente}, M., \& {Colafrancesco}, S.
  2004, \apj, 608, 166

\bibitem[{{Pfrommer} \& {En{\ss}lin}(2004)}]{pfrommer.2004}
{Pfrommer}, C. \& {En{\ss}lin}, T.~A. 2004, \aap, 413, 17

\bibitem[{{Punch} \& {Fegan}(1991)}]{punch.1991}
{Punch}, M. \& {Fegan}, D.~J. 1991, in AIP Conf. Proc. 220: High Energy Gamma
  Ray Astronomy, 321--328

\bibitem[{{Reimer} {et~al.}(2003){Reimer}, {Pohl}, {Sreekumar}, \&
  {Mattox}}]{reimer.2003}
{Reimer}, O., {Pohl}, M., {Sreekumar}, P., \& {Mattox}, J.~R. 2003, \apj, 588,
  155

\bibitem[{{Rephaeli}(1977)}]{rephaeli.1977}
{Rephaeli}, Y. 1977, \apj, 218, 323

\bibitem[{{Rephaeli} {et~al.}(1999){Rephaeli}, {Gruber}, \&
  {Blanco}}]{rephaeli.1999}
{Rephaeli}, Y., {Gruber}, D., \& {Blanco}, P. 1999, \apjl, 511, L21

\bibitem[{{Reynolds} {et~al.}(1993){Reynolds}, {Akerlof}, {Cawley}, {Chantell},
  {Fegan}, {Hillas}, {Lamb}, {Lang}, {Lawrence}, {Lewis}, {Macomb}, {Meyer},
  {Mohanty}, {O'Flaherty}, {Punch}, {Schubnell}, {Vacanti}, {Weekes}, \&
  {Whitaker}}]{reynolds.1993}
{Reynolds}, P.~T., {Akerlof}, C.~W., {Cawley}, M.~F., {Chantell}, M., {Fegan},
  D.~J., {Hillas}, A.~M., {Lamb}, R.~C., {Lang}, M.~J., {Lawrence}, M.~A.,
  {Lewis}, D.~A., {Macomb}, D., {Meyer}, D.~I., {Mohanty}, G., {O'Flaherty},
  K.~S., {Punch}, M., {Schubnell}, M.~S., {Vacanti}, G., {Weekes}, T.~C., \&
  {Whitaker}, T. 1993, \apj, 404, 206

\bibitem[{{Ryu} {et~al.}(2003){Ryu}, {Kang}, {Hallman}, \& {Jones}}]{ryu.2003}
{Ryu}, D., {Kang}, H., {Hallman}, E., \& {Jones}, T.~W. 2003, \apj, 593, 599

\bibitem[{{Sanders} {et~al.}(2005){Sanders}, {Fabian}, \&
  {Dunn}}]{sanders.2005}
{Sanders}, J.~S., {Fabian}, A.~C., \& {Dunn}, R.~J.~H. 2005, \mnras, 360, 133

\bibitem[{{Scharf} \& {Mukherjee}(2002)}]{scharf.2002}
{Scharf}, C.~A. \& {Mukherjee}, R. 2002, \apj, 580, 154

\bibitem[{{Struble} \& {Rood}(1999)}]{struble.1999}
{Struble}, M.~F. \& {Rood}, H.~J. 1999, \apjs, 125, 35

\bibitem[{{Tsuchiya} {et~al.}(2004){Tsuchiya}, {Enomoto}, {Ksenofontov},
  {Mori}, {Naito}, {Asahara}, {Bicknell}, {Clay}, {Doi}, {Edwards}, {Gunji},
  {Hara}, {Hara}, {Hattori}, {Hayashi}, {Itoh}, {Kabuki}, {Kajino}, {Katagiri},
  {Kawachi}, {Kifune}, {Kubo}, {Kurihara}, {Kurosaka}, {Kushida}, {Matsubara},
  {Miyashita}, {Mizumoto}, {Moro}, {Muraishi}, {Muraki}, {Nakase}, {Nishida},
  {Nishijima}, {Ohishi}, {Okumura}, {Patterson}, {Protheroe}, {Sakamoto},
  {Sakurazawa}, {Swaby}, {Tanimori}, {Tanimura}, {Thornton}, {Tokanai},
  {Uchida}, {Watanabe}, {Yamaoka}, {Yanagita}, {Yoshida}, \&
  {Yoshikoshi}}]{tsuchiya.2004}
{Tsuchiya}, K., {Enomoto}, R., {Ksenofontov}, L.~T., {Mori}, M., {Naito}, T.,
  {Asahara}, A., {Bicknell}, G.~V., {Clay}, R.~W., {Doi}, Y., {Edwards}, P.~G.,
  {Gunji}, S., {Hara}, S., {Hara}, T., {Hattori}, T., {Hayashi}, S., {Itoh},
  C., {Kabuki}, S., {Kajino}, F., {Katagiri}, H., {Kawachi}, A., {Kifune}, T.,
  {Kubo}, H., {Kurihara}, T., {Kurosaka}, R., {Kushida}, J., {Matsubara}, Y.,
  {Miyashita}, Y., {Mizumoto}, Y., {Moro}, H., {Muraishi}, H., {Muraki}, Y.,
  {Nakase}, T., {Nishida}, D., {Nishijima}, K., {Ohishi}, M., {Okumura}, K.,
  {Patterson}, J.~R., {Protheroe}, R.~J., {Sakamoto}, N., {Sakurazawa}, K.,
  {Swaby}, D.~L., {Tanimori}, T., {Tanimura}, H., {Thornton}, G., {Tokanai},
  F., {Uchida}, T., {Watanabe}, S., {Yamaoka}, T., {Yanagita}, S., {Yoshida},
  T., \& {Yoshikoshi}, T. 2004, \apjl, 606, L115

\bibitem[{{V{\" o}lk} \& {Atoyan}(1999)}]{voelk.1999}
{V{\" o}lk}, H.~J. \& {Atoyan}, A.~M. 1999, Astroparticle Physics, 11, 73

\bibitem[{{Weekes}(1996)}]{weekes.1996}
{Weekes}, T.~C. 1996, Space Science Reviews, 75, 1

\end{thebibliography}
\bibliographystyle{apj}
\clearpage
\begin{table}
  \caption{\label{tab:datasets} Description of the various data sets
    used in this analysis.  The Crab sets titled ``Crab-0.5'' and
    ``Crab-0.8'' are observations performed with the telescope offset
    from the location of the Crab nebula by 0.5$^\circ$ and 0.8$^\circ$,
    respectively.}
\begin{center}
\begin{tabular}{||c|c|c|c|c||}\hline \hline
  Source & Season & Number &  ON   & OFF   \\
  ~   &   (MJD)    & (Pairs) & (min) & (min) \\        
  \hline \hline
  Perseus & 2004-2005 & 29 & 810.4 & 810.4 \\ 
  \hline
  Abell 2029   & 2003-2004 & 13 & 363.3 & 363.3 \\
  \hline
  Crab    & 2003-2004 & 29 & 810.4 & 810.4 \\
  \hline
  Crab    & 2004-2005 & 24 & 670.7 & 670.4 \\
  \hline
  Crab-0.5& 2003-2005 & 6 & 167.7 & 167.6 \\
  \hline
  Crab-0.8& 2003-2005 & 8 & 223.6 & 223.6 \\
  \hline
\end{tabular}
\end{center}
\end{table}

\begin{table}
\caption{\label{tab:king} Values of the double-$\beta$ model parameters
  for the Perseus cluster of galaxies from \cite{pfrommer.2004} and
  based on data from \cite{churazov.2003} and \cite{struble.1999}.
  The values shown for Abell 2029 are for a King profile based on data
  from \cite{jones.1984}.}
\begin{center}
\begin{tabular}{||c|c|c|c|c|c|c||}\hline \hline
Cluster & $a_1$ & $a_2$ & ~$r_1$~ & ~$r_2$~ & ~$\beta_1$~ & ~$\beta_2$~\\
~ &~&~&(kpc)&(kpc)& & \\        
\hline \hline
Perseus & 1.0 & 0.104 & 57 & 200 & 1.2 & 0.58 \\
\hline
Abell 2029   & 1.0 & N/A   &212 & N/A & 0.83 & N/A \\
\hline
\end{tabular}
\end{center}
\end{table}

\begin{table}   
\caption{\label{radio} Gamma-ray flux 90\% upper limits on
    spectroscopically resolved radio galaxies associated with members
    of the Perseus and Abell 2029 clusters of galaxies.  The 20 cm
    flux data are from The NRAO VLA Sky Survey \citep{condon.1998}.}
\begin{center}
\begin{tabular}{||c|c|c|c|c|c|c||} \hline \hline

~Cluster~ & ~Source~  &   ~RA~  &  ~DEC~  & ~20 cm Flux~ &\multicolumn{2}{c||}{~400 GeV Flux Upper Limit~} \\ 
          &           & (J2000) & (J2000) &    (mJy)    & (Crab) & ($10^{-11}$ergs~cm$^{-2}$~s$^{-1}$) \\
                             \hline \hline

Perseus & 3C 84.0 & 03 19 48 & +41 30 42 & 2829.2 & 0.047 & 0.29\\
& (NGC 1275) & & & & & \\
\hline
Perseus & 3C 83.1 & 03 18 16 & +41 51 27 & 1305.5 & 0.086 & 0.53\\
& (NGC 1265) & & & & & \\
\hline
Perseus & IC 310     & 03 16 43 & +41 19 29 & 168.1  & 0.13  & 0.80\\
\hline
Abell 2029 & IC 1101 & 15 10 56 & +05 44 42 & 527.8  & 0.13  & 1.1\\
\hline
\end{tabular}
\end{center}
\end{table}

\begin{table}
  \caption{\label{limits} Upper limits for the diffuse CRp emission from Perseus and
    Abell 2029 using various angular cuts.  The 0.2$^\circ$ cut is relevant for
    point source and dark matter emission.  The 0.3$^\circ$ cut is the
    optimal cut for the extended emission while the 0.8$^\circ$ one
    contains the emission from a large fraction of the
    field-of-view. The scaling factor is used to convert upper limits
    from Crab units to differential fluxes, taking into account the
    expected spectral shape.}
\begin{center}
\begin{tabular}{||c|c|c|c|c|c||}\hline \hline
Cluster & Angular Cut  & Significance & \multicolumn{2}{c|}{400~GeV
  Flux Upper Limit} & Scaling Factor\\
~ & (Degrees) & (Sigma) & (Crab) & ($10^{-11}$ergs~cm$^{-2}$~s$^{-1}$)
& \\        
\hline \hline
Perseus & 0.2 & -2.3 & 0.047 & 0.29 & 0.80\\
\hline
Perseus & 0.3 & -2.1 & 0.13  & 0.80 & 0.80\\
\hline
Perseus & 0.8 & -0.91 & 0.12  & 0.74 & 0.80\\
\hline
Abell 2029 & 0.2 & -1.2 & 0.10 & 0.87 & 1.1 \\
\hline
Abell 2029 & 0.3 & -0.15 & 0.14 & 1.2 & 1.1\\
\hline
Abell 2029 & 0.8 & -0.79 & 0.25 & 2.2 & 1.1\\
\hline
\end{tabular}
\end{center}
\end{table}

\clearpage

\end{document}